\documentclass[%
 reprint,
 amsmath,amssymb,
 aps,
showkeys
,longbibliography
]{revtex4-2}

\usepackage{graphicx}
\usepackage{dcolumn}
\usepackage{bm}
\usepackage{physics}
\usepackage{placeins}
\usepackage{subscript}
\usepackage[super]{nth}
\usepackage{caption}
\usepackage{afterpage}
\usepackage{xcolor}  
\usepackage{cleveref}

\usepackage{subcaption}
\usepackage{float}

\usepackage[font=small,labelfont=bf]{caption}

\begin{document}

\preprint{APS/123-QED}

\author{Adam Herling}
\affiliation{%
  Technion- Israel Institute of Technology, Schulich Faculty of Chemistry and Faculty of Physics, Haifa, 32000036, Israel
}%

\author{Ofer Neufeld}%
 \email{ofern@technion.ac.il}
\affiliation{%
  Technion- Israel Institute of Technology, Schulich Faculty of Chemistry, Haifa, 32000036, Israel
}%


\title{Role of ultrafast electron-optical-phonon interactions in high harmonic generation from graphene
}

\begin{abstract}
High harmonic generation (HHG) is a commonly explored process across material systems, where intense lasers drive attosecond-to-femtosecond electron dynamics within solid bands, causing high-energy emission. The main physical players in HHG are the electrons and photons, who are commonly thought to dictate the HHG spectral properties. However, solids also host ubiquitous phonons that are usually relevant on longer timescales, and are therefore largely assumed negligible in HHG. In general, it is unclear if/how phonons partake in HHG and in dephasing of the electron dynamics, which has been very recently proposed in different contexts. We theocratically study HHG in graphene with a formalism that includes optical phonons in the static limit, where the lattice is frozen on the electronic timescale and HHG is computed by sampling thermally-occupied phonons and ensemble-averaging. We show that in graphene: (i) Optical-phonons strongly suppress HHG yields by coupling to interband currents and causing harmonic phase scrambling (inducing destructive interference). This explains lack of experimental observation of HHG above $\sim$3 eV from graphene. (ii) HHG yields become temperature-dependent due to phonon occupations, though in graphene this dependence is weak since phonon energy scales are dominated by zero-point motion. (iii) Optical phonons dephase interband coherences in a rate roughly equivalent to $T_2\sim 5.7$ fs. This timescale is substantially faster compared to $e-e$ scattering, suggesting that thermal phonons dominate electronic decoherence in strong-fields. (iv) Phonons smoothen HHG ellipticity-dependent curves, yielding better matches with experiments. Remarkably, all of these effects are timescale-independent, as they arise in the static picture of electron-phonon interactions, meaning results are transferable to attosecond phenomena. Our results shed light on the dephasing time problem in HHG and role of phonons on attosecond timescales, and should be transferable to other systems and processes as well (e.g. Floquet gaps and photocurrents in graphene), motivating novel spectroscopies of phonon dynamics. 

\end{abstract}


\maketitle


Since it's seminal observation in 2011\cite{Shambhu2011NaturePhys}, high harmonic generation (HHG) in solids has been heavily studied both theocratically and experimentally\cite{Ghimire2019,GaardeTutorialHHGSolids2021}. One main motivator for the field is the potential application of HHG for probing material properties in- and out-of-equilibrium. Since the main HHG mechanism arises from a combination of intraband emission and interband coherences\cite{Wu2016, GaardeTutorialHHGSolids2021}, the process is directly sensitive to the band structure and curvature, as well as to electronic ultrafast band occupation dynamics. This has been applied for a variety of spectroscopies, including reconstructions of bands\cite{PhysRevLett.115.193603,Lanin2017,Lv2021}, Berry curvatures\cite{Luu2018b,Lv2021,Uzan-Narovlansky2024}, coherent phonon dynamics\cite{Bionta2021,Neufeld2022g,DixitHHGGraphenePRA2022,Zhang2024,Zhang2024a}, valley occupations\cite{Jimenez-Galan2020,Mitra2024,Tyulnev2024,Lively2024,Ofer2025SolidsNatCommun}, etc. Recent reports also connect HHG emission to Floquet band dressing\cite{Uzan-Narovlansky2022,Neufeld2022d,Galler2023,Mitra2024,zhao2025high}. 

One main gap in current theory of HHG is that it is understood purely at the electronic level, with electrons interacting only with laser photons (classical or quantum\cite{Gothelf2025,Lange2025}). Several works in recent years also allow electron-electron ($e-e$) interactions at various levels\cite{PhysRevLett.121.097402,Silva2018a,OferPRX2023,Valmispild2024,Jensen2024a,ChangLee2024,Molinero2024}, but broadly speaking, consideration of other particles in HHG mechanisms has been limited. This is in contradiction to modern theory of solid-state, where phonons are ubiquitous and omnipresent, having dominant contributions to various phenomena from heat capacity to superconductivity. Most recently, coherently-pumped phonons were probed in HHG\cite{Bionta2021,Neufeld2022g,DixitHHGGraphenePRA2022,Zhang2024,Zhang2024a}, where electron-phonon interactions were included. However, in these works the phononic coherence was essential, and the usual regime where phonons are populated thermally and incoherently is poorly understood. This open question is connected to another major challenge - the problem of HHG ultrafast dephasing times ($T_2$). It has been established across multiple works that extremely short phenomenological dephasing times (down to few femtoseconds or less) need to be used in simulations to obtain `clean' spectra matching experiments (e.g. see refs. \cite{Brown:24,Kolesik2023a,Wang2021a,Korolev2024}). These timescales grossly mismatch with expectations based on separate measurements of coherence times that often yield many tens or hundreds of femtoseconds\cite{Mai2014,Wang2018}, even in strong fields\cite{Heide2021,Lively2024,Mitra2024}. Several different works proposed various solutions to this conundrum, including propagation and macroscopic physics\cite{Floss2019}, quantum-optical effects\cite{Bae2026}, dynamical electron-phonon coupling\cite{Freeman2022}, and Brillouin-zone (BZ) averaged dephasing due to electron-phonon interactions\cite{Korolev2024,Luo2026}. Generally, the idea that phonons might be in-charge of ultrafast dephasing has been slowly taking hold, including an earlier work that connected dephasing with temperature effects in toy models\cite{Du2022} (though there HHG yields were not greatly impacted and mostly the cutoff was explored). Still, the mechanism that allows phonons to act on such fast timescales remains vague. BZ-averaging might lead to faster dephasing in some light-matter regimes\cite{Korolev2024}, but in large-gap systems that mechanism is unlikely. It is also not clear if optical or acoustic phonons are in charge\cite{Freeman2022}. We further note that while working on this manuscript, we became aware of three other recent reports that deal with similar questions but from slightly different angles\cite{cardenas2025effects,mokhtari2026phonon,hatch2026probinglatticefluctuationsusing}. 

Here we theocratically explore HHG from graphene, a prototypical 2D material used in many HHG and ultrafast experiments\cite{Yoshikawa2017,Hafez2018,Baudisch2018,Cha2022,Chen2025a}. We simulate HHG using two-band semiconductor Bloch equations (SBE) coupled to graphene optical in-plane phonon modes (transverse and longitudinal). We thermally occupy these modes and sample their quantum distribution, obtaining the laser-driven current via ensemble averaging (as arises in statistical mechanics approaches such as those recently employed in liquid HHG\cite{PhysRevLett.124.203901,Xu2025,Mondal2025}). This scheme effectively mimics the experimental conditions where HHG is contributed by emission from large crystal volumes with numerous unit cells with random phononic phases. Using this technique, we study HHG temperature dependence, ultrafast dephasing, and dependence on the laser parameters. We find that HHG from graphene is strongly suppressed above $\sim3$ eV. The suppression is largely temperature-independent due the graphene phononic energy scales, but can lead to temperature dependence in other solids depending on their phononic bands. This result explains lack of experimental observations of harmonics above $\sim3.1$ eV\cite{Yoshikawa2017,Hafez2018,Baudisch2018,Cha2022,Chen2025a} in graphene, despite numerous simulations consistently predicting higher-energy emission. We study the mechanism behind this effect and show that it arises from optical phonons `scrambling' HHG phases, inducing strong destructive interferences (while perturbative harmonics remain phase-synced). This effect arises only in the interband emission channel. This result is remarkable considering that the timescales of phononic motion are completely irrelevant - the static lattice displacements cause phase scrambling in a timescale-independent mechanism that should also appear in attosecond experiments. We further analyze the dephasing dynamics of the interband coherences, finding decays equivalent to $T_2\sim 5.7$ fs, indicating that in HHG and strong-field physics electron-phonon scattering is the dominant decoherence channel rather than $e-e$ scattering. Lastly, we show that optical phonons can modulate the HHG ellipticity dependence, posing an interesting channel for probing temperature-dependent effects and phononic occupations on ultrafast timescales.

Let us begin by describing our methodological approach and employed formalism, starting with the standard phonon-free case. In absence of interactions with phononic degrees of freedom (DOF), electrons in graphene are described by a second-order nearest-neighbor tight binding (TB) Hamiltonian of the following form:
\begin{equation}
    H_{0}(\textbf{k}) = 
\begin{bmatrix}
    t_2\sum_{j} {e^{-i\textbf{v}_{2,j}\cdot\textbf{k}}} & t_1\sum_{j} {e^{-i\textbf{v}_{1,j}\cdot\textbf{k}}} \\
    t_1\sum_{j} {e^{i\textbf{v}_{1,j}\cdot\textbf{k}}} & t_2\sum_{j} {e^{-i\textbf{v}_{2,j}\cdot\textbf{k}}}
\end{bmatrix}
\end{equation}
where $t_{1,2}$ are first and second neighbor hopping terms, respectively, $\textbf{v}_{1,j}$ and $\textbf{v}_{2,j}$ are the first and second neighbor connecting vectors, with the sums running over all neighbors of each kind, and $\textbf{k}$ is the Bloch momenta. $H_0$ is diagonalized analytically, from which we obtain the energy bands $\epsilon_{CB/VB}(\textbf{k})$, and the eigenstates $\ket{u_{CB/VB}(\textbf{k})}$, of the valence (VB) and conduction bands (CB). From $\ket{u_{CB/VB}(\textbf{k})}$ the transition dipole ($\textbf{d}_{nm}^\textbf{k}=i\bra{u_{n}(\textbf{k})}\partial_{\textbf{k}}\ket{u_{m}(\textbf{k})}$) and momentum matrix elements ($\textbf{p}^\textbf{k}_{nm}=\bra{u_{n}(\textbf{k})}\partial_{\textbf{k}}H_0(\textbf{k})\ket{u_{m}(\textbf{k})}$) are analytically evaluated (similar to the approach employed in refs. \cite{Ofer2025SolidsNatCommun,Chen2025a}, with the gauge choice specified in ref. \cite{Neufeld2026}).

The interaction of an intense femtosecond laser pulse with graphene electrons is captured within the SBE in the length gauge, Houston basis, and while applying the dipole approximation. These are given in atomic units by\cite{GaardeTutorialHHGSolids2021}:
\begin{equation}
\begin{aligned}
    i \frac{d}{dt} \rho^{\mathbf{k}}_{mn} =& 
    ( \epsilon_{m}(\mathbf{k}(t)) - \epsilon_{n}(\mathbf{k}(t))) \rho^{\mathbf{k}}_{mn}
    + i \frac{1 - \delta_{mn}}{T_2} \rho^{\mathbf{k}}_{mn} \\
    &- \mathbf{F}(t) \cdot \sum_{l} [
    \mathbf{d}^{\mathbf{k}(t)}_{ml} \rho^{\mathbf{k}}_{ln} 
    - \mathbf{d}^{\mathbf{k}(t)}_{ln} \rho^{\mathbf{k}}_{ml}
    ],
\end{aligned}
\end{equation}
where $\rho^{\mathbf{k}}_{mn}$ is the density matrix term connecting bands $n$ and $m$ at \textit{k}-point \textbf{k}. $\mathbf{F}(t)$ and $\mathbf{A}(t)$ are the electric field and vector potential of the driving laser, respectively, which are related via $c\mathbf{F}(t) = - d \mathbf{A}(t) / dt$. Throughout the text we employ a generic elliptically-polarized laser pulse of the form $\textbf{A}(t)=\frac{f(t) E_0}{\omega c \sqrt{1+\varepsilon^2}}[\cos(\omega t) \hat{\textbf{x}} +\varepsilon \sin(\omega t)\hat{\textbf{y}}]$, with $E_0$ the field amplitude, $c$ the speed of light, $\omega$ the laser frequency, $\varepsilon$ the laser ellipticity, and $f(t)$ a temporal envelope of duration $T_{pulse}=8\frac{2\pi}{\omega}$, yielding a full-width-half-max of $\sim 35$ fs (see SI for details). We apply the Peierls substitution in the Houston gauge, such that $\textbf{k}(t)=\textbf{k}+\textbf{A}(t)/c$. A phenomenological term accounting for decoherence can be added with a constant dephasing time $T_2$ regardless of the inclusion of phononic interactions that will be next described. 

These equations are solved numerically (see SI), where at each time step we compute the time-dependent current:
\begin{equation}
    \mathbf{J}(t) 
    = \sum_{m,n,\textbf{k}}   w_{\textbf{k}} \mathbf{p}^{\mathbf{k}(t)}_{mn}   \rho^{\mathbf{k}}_{mn}
\end{equation}
, with $w_{\textbf{k}}$ the \textit{k}-point weight. Note that $\textbf{J}(t)$ can be separated to intraband terms ($n=m$ case) or interband ($n\neq m$) terms\cite{Yue2022a}. From the currents we obtain the HHG emission as the spectral power of the Fourier transform of $\partial_t\textbf{J}(t)$. 

\begin{figure}[h]
    \centering
    \includegraphics[width=0.75\columnwidth]{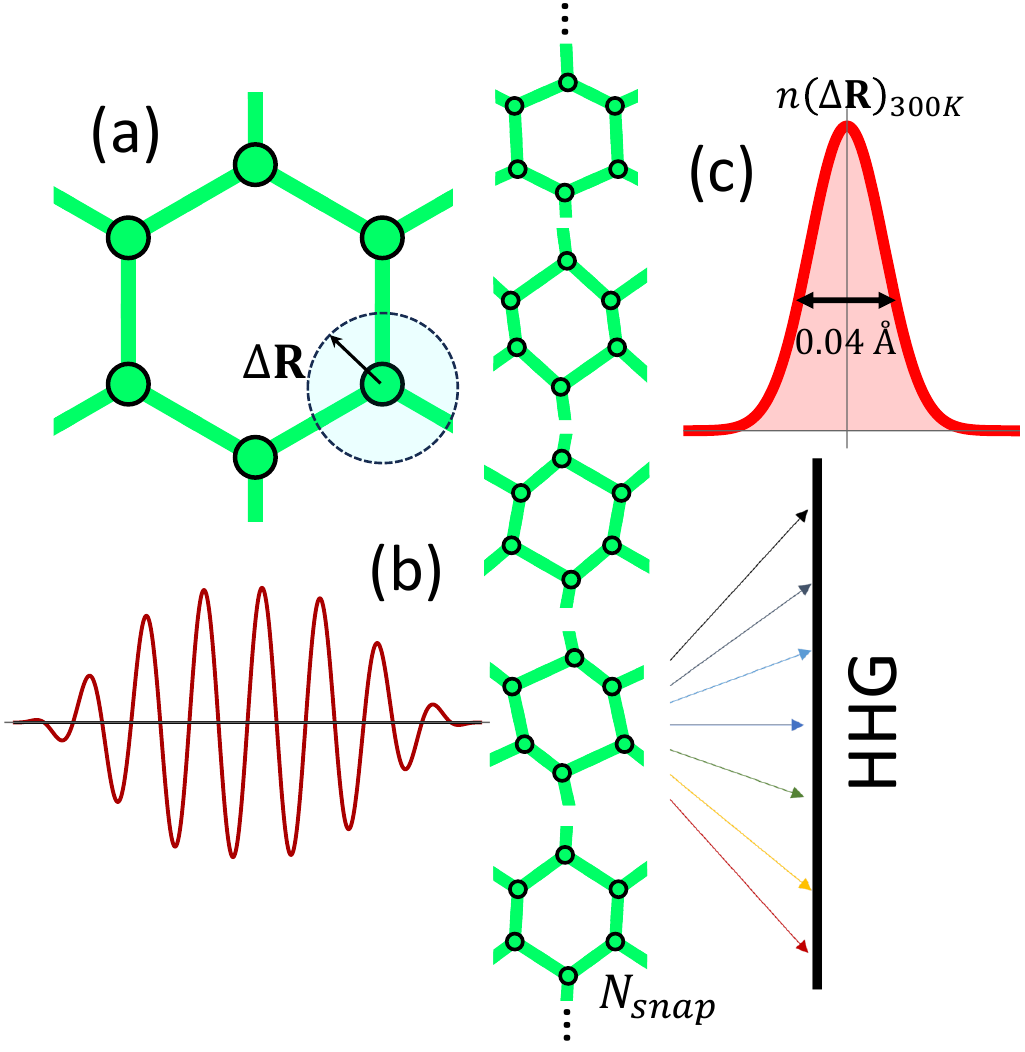}
    \captionsetup{margin=0pt}
    \caption{Illustration of thermal optical phonons in graphene at $\Gamma$. The graphene lattice (including phonons) is driven by an intense laser, causing HHG. (a) Equilibrium lattice ($\Delta \textbf{R}=0$). Possible $\Delta \textbf{R}$ displacements are indicated by the arrow and circle around one atom. (b) Series of snapshots for varying phonon-induced lattice displacements, generating static lattice configurations that are sampled with proper weights to include phononic interactions. The displacements are grossly exaggerated for illustrative purposes. (c) Exemplary phononic displacement distribution function at 300K.}
    \label{fig:1}
\end{figure}

To contrast with the phonon-free equilibrium lattice case, we also describe HHG including phononic DOF by thermally populating the longitudinal optical (LO) and transverse optical (TO) phonon modes at $\Gamma$. 
The effects of these modes on $H_0$ is twofold. First, by populating a phonon mode the lattice geometry slightly distorts (see illustration in Fig. \ref{fig:1}), which is described by shifting one of the two atoms in the unit cell positions away from equilibrium by $\Delta \textbf{R}$, while fixing the lattice vectors. This alters the TB nearest-neighbor vectors ($\textbf{v}_{1,j}$). Second, the shifted lattice geometry affectively alters the hopping terms to different neighbors (with closer neighbors permitting larger hopping and vice versa). We model the changes in hopping coefficients with an exponential decaying function fitted to graphene parameters\cite{Ribeiro_2009}, $\tilde{t_1}(\Delta \textbf{R})$ (see SI). The resulting Hamiltonian with displacement reads:
\begin{equation}
    H^{\Delta \textbf{R}}_{0}(\textbf{k}) =
    \begin{bmatrix}
    t_2\sum_{j} {e^{-i{{\textbf{v}}}_{2,j}\cdot\textbf{k}}} & \tilde{t_1}(\Delta \textbf{R})\sum_{j} {e^{-i\tilde{\textbf{v}}_{1,j}\cdot\textbf{k}}} \\
    \tilde{t_1}(\Delta \textbf{R})\sum_{j} {e^{i\tilde{\textbf{v}}_{1,j}\cdot\textbf{k}}} & t_2\sum_{j} {e^{-i{\textbf{v}}_{2,j}\cdot\textbf{k}}}
\end{bmatrix}
\end{equation}
where $\tilde{\textbf{v}}_{1,j}$ are the nearest-neighbor vectors including displacements that are no longer of equal length. Note that second-order hopping amplitudes and vectors remain unchanged with $\Gamma$-only phonons since they are A-A and B-B sublattice connections. Since $H^{\Delta \textbf{R}}_{0}$ is still $2\times2$, we analytically diagonalize it and obtain exact equations for the bands, states, and dipole and momentum matrix elements, which become functions of $\Delta \textbf{R}$. The same SBE formalism as in the equilibrium case is applied, where the value $\Delta \textbf{R}$ is sampled from the phononic distribution and fixed during the simulation (i.e. $\partial_t\Delta \textbf{R}=0$). We refer to this as the static phonon approximation\cite{Lively2024}, which should be reasonable on ultrafast timescales. In other words, electron-phonon interactions are included only in a forward direction, with phonons acting on electrons, but not the other way around. The temperature determines the width of the distribution function, calculated per phonon polarization from the variance of the position operator in a corresponding harmonic oscillator $\sigma^2(T)=\frac{\hbar}{M_c\omega_{ph}}\coth(\frac{\hbar\omega_{ph}}{2k_BT})$, with $M_c$ the mass of a carbon atom and $\omega_{ph}$ the frequency of the $\Gamma$ phonon. For instance, at 300K we obtain a typical value for graphene of $STD(\Delta \textbf{R})|_{300K}=0.0417$ \AA, amounting to $\sim 1.7$\% of the lattice parameter (i.e. rather small displacements). Separate and independent simulations are performed for many values of $\Delta \textbf{R}$ that sample static `snapshots' of the lattice, and the current from these individual simulations is coherently summed to obtain the total current that includes phonon contributions (much like recent schemes in liquid HHG\cite{Xu2025,Mondal2025}). The simulation is formally converged with the number of snapshots ($N_{snap}$, see data in SI). Note that besides the phononic interactions, the $T_2$ term is still applicable and can be tested for additional effects in HHG that might be attributed to $e-e$ scattering or other phononic channels beyond optical, and beyond $\Gamma$.  

\begin{figure*}[t!]
    \centering
    \includegraphics[width=0.99\textwidth]{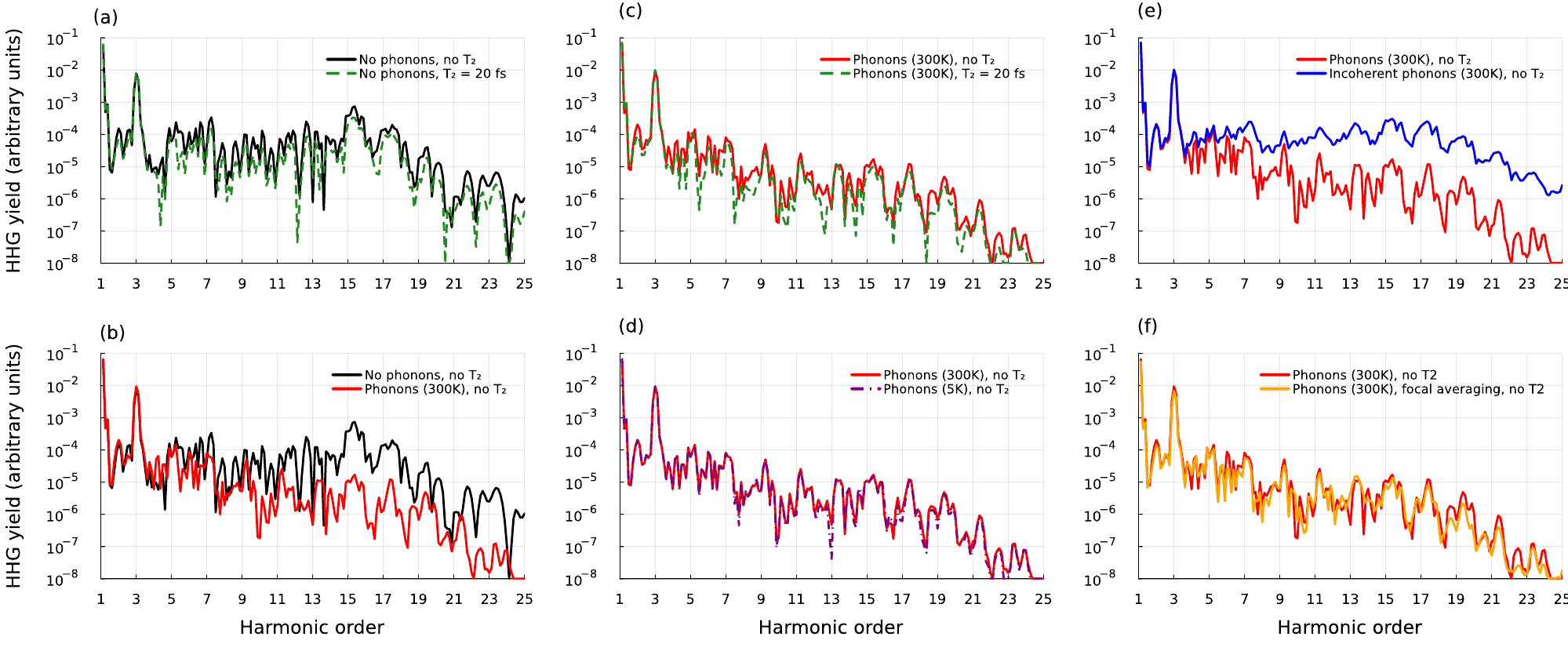}
    \captionsetup{margin=0pt}

    \caption{HHG spectra from graphene simulated at varying levels of theory. (a) HHG at the equilibrium case (no phononic DOF) with/without phenomenological dephasing. (b) HHG emission with phononic DOF compared to equilibrium case, showing strong suppression of the plateau and spectral cleaning. (c) HHG emission including phononic DOF and with/without phenomenological dephasing. (d) Temperature-dependence of HHG in graphene calculated including phononic DOF and without phenomenological dephasing. (e) Comparison of coherent/incoherent summations over HHG phononic snapshots. (f) HHG emission including phononic DOF with/without Gaussian focal beam averaging. Calculations performed for a peak laser power of $1.15\times10^{12}$ W/cm$^2$ and wavelength of 2600 nm. Data presented in logarithmic scale.}
    \label{fig:2}
\end{figure*}

We now employ these two approaches to study HHG in graphene. Figure \ref{fig:2} presents HHG driven by linearly-polarized pulses at typical experimental conditions ($\lambda=2600$ nm, $I_0=1.15\times10^{12}$ W/cm$^2$). We observe, consistently with many prior works, that for $\Delta \textbf{R}=0$ (denoted as the equilibrium case), a wide HHG plateau is obtained up to $\sim 8$ eV. This result is apparent even when including phenomenological dephasing at a reasonable timescale of $T_2=20$ fs\cite{Heide2021} (see Fig. \ref{fig:2}(a)), which only cleans-up the harmonic spectra and symmetrizes harmonic peaks, as expected (in absence of phenomenological dephasing the spectrum is extremely noisy). On the other hand, inclusion of optical phonons (Fig. \ref{fig:2}(b)) substantially suppresses the HHG yields of non-perturbative harmonics, such that above $\sim 3$ eV HHG is reduced by an order of magnitude or more. This effect is in agreement with multiple experiments that to date could not resolve high harmonics above $\sim 3.1$ eV from graphene. Therefore, our theory suggests that lack of high harmonic emission from graphene is a result of optical-phonon induced suppression. 

Before exploring the physical origin of this effect, we further study its Temperature dependence and find that HHG is largely temperature-independent (Fig. \ref{fig:2}(d)). Physically, this occurs in graphene due to optical phonons energy scales - At $\Gamma$ phonons arise at $\sim200 meV$\cite{Piscanec2007}, such that at room temperature phonons are only $\sim0.2\%$ occupied. Thus, minor occupations dominated by zero point quantum-nuclei motion lead to the strong HHG suppression, and temperature reduction hardly changes the HHG emission. It is worth noting that under intense laser driving, it is very likely that much higher phonon occupations of optical modes occurs due to indirect heating. Generally, we notice a substantial shift in HHG yields occurring only towards $\sim1500K$ due to these energy scales (see SI).

We next explore the role of phenomenological dephasing introduced through the $T_2$ term and the level of `cleanliness' of harmonic peaks. In simulations including phononic DOF a $T_2$ term can also be added, which might account for $e-e$ scattering channels, or other acoustic and non-$\Gamma$ phononic interactions. We generally observe in simulations that HHG spectra including phononic DOF but without $T_2$ dephasing is also substantially `cleaned', where HHG peaks are symmetrized and noise in-between harmonics is reduced. This is somewhat analogous to HHG with addition of $T_2$, though there still exists minor noise in-between harmonics and some asymmetric harmonic profiles observed. Figure \ref{fig:2}(c) presents HHG spectra including both phononic DOF and a $T_2$ term, which adds an additional minor cleaning effect, just as in the equilibrium case. In that respect, added phenomenological dephasing plays the same role even if it is combined with optical-phonon-electron interactions in the static approximation. Overall, we conclude that optical phonons in graphene have a similar impact to phenomenological dephasing in cleaning the spectrum.

Next we consider other potential origins for peak cleaning features observed in experiments. Figure \ref{fig:2}(e) presents a comparison of coherent and incoherent summation over HHG spectra from the various snapshots employed in the phononic case (with $T_2\rightarrow\infty$). The correct physical procedure involves a coherent sum, as employed throughout. Nonetheless, we observe that the incoherent sum has an extremely `cleaning' effect on HHG spectra (see Fig. \ref{fig:2}(e)), while the yield suppression vanishes as expected. This motivates us to explore macroscopic beam focal averaging. HHG yields evaluated including focal averaging require the various phononic snapshot HHG spectra to be summed at two levels, both coherently within a given microscopic region, as well as incoherently from regions in the gaussian beam that are distant from each other on length scales beyond electronic coherence\cite{de2025fully}, each with proper weights (see details in SI). Figure \ref{fig:2}(f) presents this procedure, showing that both the suppression, and slightly cleaner symmetrized HHG peaks, survive (though the effect is rather small and mostly noticeable in-between harmonic peaks).
\begin{figure*}[t!]
    \centering
    \includegraphics[width=0.95\textwidth]{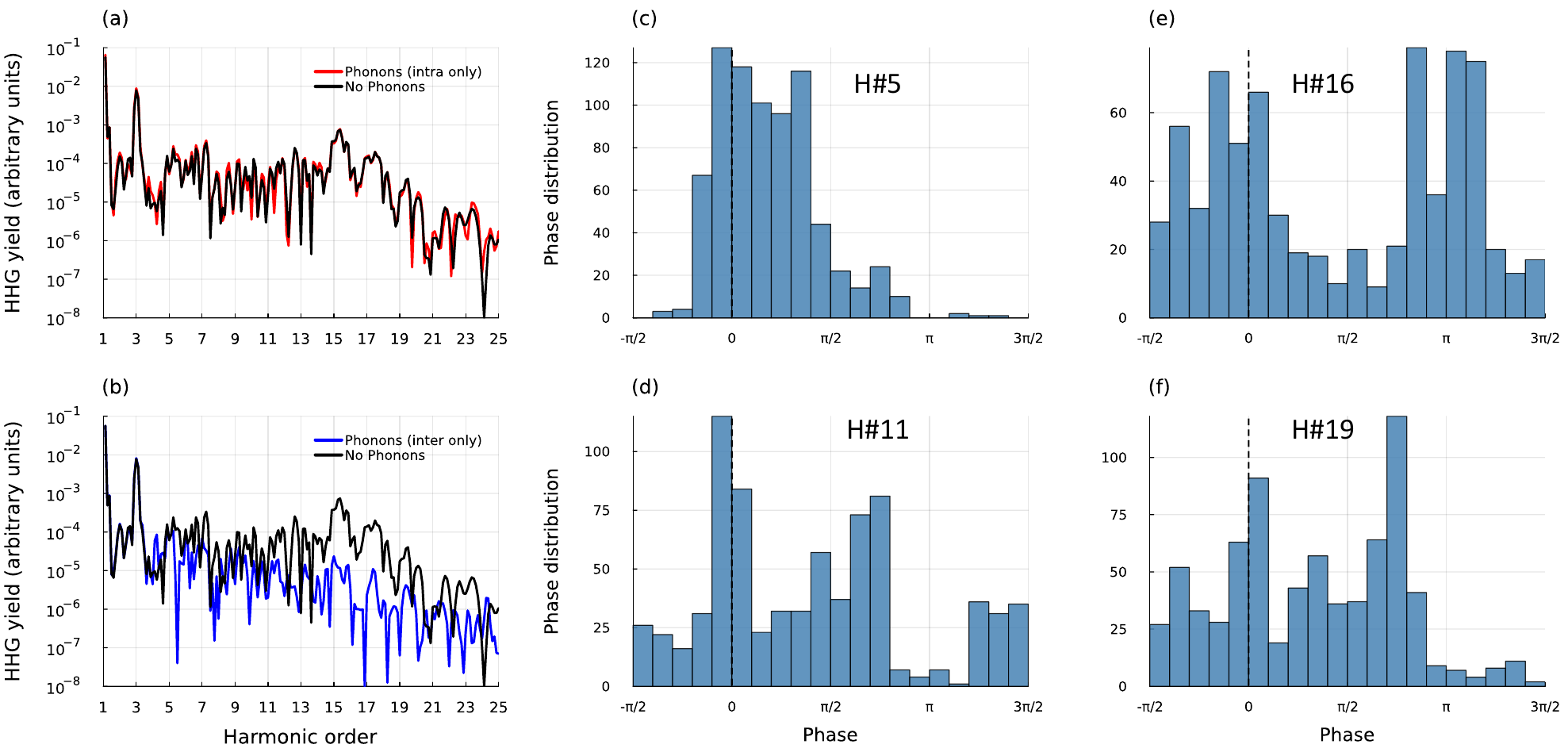}
    \captionsetup{margin=0pt}

    \caption{(a) HHG emission with phononic DOF included only in intraband channel, compared to the equilibrium case. (b) Same as (a) but for interband emission. (c-f) Select harmonic order phase distributions across phononic snapshots from the interband emission channel (all other harmonics are delegated to the SI). Simulations performed in similar conditions to Fig. 2 without phenomenological dephasing. Zero phase in each case is offset to the equilibrium case phase (dashed black line).}
    \label{fig:3}
\end{figure*}

Having established prominent effects of optical phonons in HHG from graphene, we turn to analyze their physical origin. At a first step, we separate the phononic contribution to HHG emission in the inter/intraband channels. Figure \ref{fig:3}(a) shows HHG spectra computed by including phonon DOF in the intraband emission, but fixing the interband emission to the equilibrium case, which isolates the impact of phonons on intraband harmonics. The results clearly show that interaction between phonon DOF and intraband currents does not induce a suppression. Contrarily, the same analysis performed by fixing the intraband currents and phonon-averaging over interband emission strongly suppresses HHG, but only for non-perturbative harmonics (largely above H5, see Fig. \ref{fig:3}(b)). This connects the effects to interband coherence of the electronic system. To understand why interband emission is so strongly suppressed, yet unaffected in the perturbative harmonics, we analyze HHG emission phases at a snapshot- and harmonic-resolved level. From each of the snapshots that comprise the interband phononic HHG case, we extract the harmonic emission phase, plotted as a statistical distribution in Fig. \ref{fig:2}(c-f) for select harmonics (see SI for additional data). In perturbative harmonics the phase distribution is narrow, suggesting minimal destructive interference (e.g. H5 in \ref{fig:3}(c)). In suppressed higher harmonics the phase distribution is very wide. For instance, H11 in Fig. \ref{fig:3}(d) shows very broad phase distribution that leads to massive destructive interference. In cases where specific phase values contribute dominantly, a counter peak at the opposite phase value arises, which also causes destructive interferences (e.g. H19 Fig. \ref{fig:3}(f)). The in-between harmonic region such as H16 that is symmetry-forbidden\cite{Neufeld2019} also exhibits this effect (Fig. \ref{fig:2}(e)), which allows spectral cleaning in-between orders. We coin this effect `phononic phase scrambling', which we expect to arise generally for sufficiently occupied phonon branches. The case of graphene is unique in that here dominantly zero-point motion is sufficient to observe strong suppression, even down to few Kelvin temperatures and very small $STD(\Delta \textbf{R})$. We hypothesize that this has to do with graphene's Berry phase around the Dirac cones that carry $\pm\pi$ values\cite{Zhang2005a,Dutreix2019}. Essentially, optical phonons in graphene do not break inversion symmetry, and therefore do not open the gap or lift the nonzero Berry phase. However, even small shifts in $\Delta \textbf{R}$ break the six-fold rotational symmetry and shifts the Dirac cones positions in \textit{k}-space\cite{Gui2008,Cocco2010}. We expect this causes an ambiguity of $\pm\phi$ in HHG phases depending on if $\Delta \textbf{R}$ is positive or negative (where both branches are roughly equally populated in the harmonic phonon approximation). It remains unclear at this stage if such a mechanism is also relevant in gapped hexagonal solids, which should be topic of future work. 

\begin{figure*}[t]
   \centering
    \includegraphics[width=0.95\textwidth]{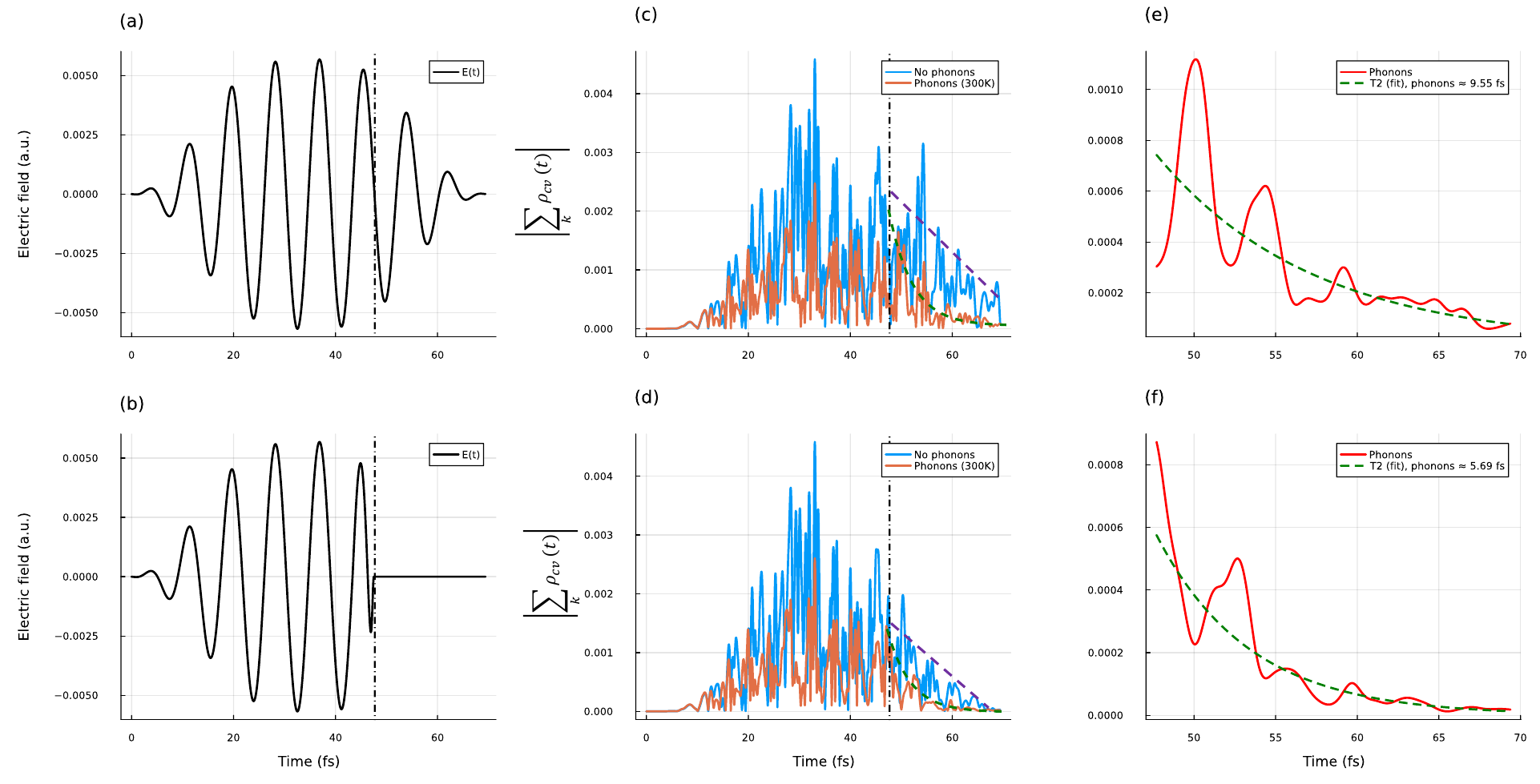}
    \captionsetup{margin=0pt}

    \caption{Interband coherence dynamics for BZ-averaged $\rho_{CV}(t)$. (a) Laser electric field. (b) Interband coherence dynamics with/without phonons. Vertical dashed line represents the temporal region towards the end of the pulse from which on the dynamics are fitted to an exponentially decaying function (dashed green). Linear dashed line in purple marks the non-exponential coherence leakage in the equilibrium case. (c) Same as (b) but in the temporal fitting region with the optimally-fitted exponential function. (d-e) Same as (a-c) but for a case where the laser electric field is artificially halted in simulations in order to obtain a cleaner extraction of dephasing. Simulations do not include phenomenological dephasing and performed in similar conditions to Fig. \ref{fig:3}.}
    \label{fig:4}
\end{figure*}

We next directly explore ultrafast interband coherence dynamics. Figure \ref{fig:4} presents BZ-averaged values for $|\sum_{\textbf{k}} \rho^{(\textbf{k})}_{cv}(t)|$, which can be explored with/without phonon DOF (phenomenological dephasing is not considered at this stage as it leads to the expected exponential decay of $\rho_{cv}(t)$ and disrupts our analysis of phonon-induced dephasing). The summation of interband coherences across the BZ is a direct part of the total interband electric current in the system (eq. 3). Indeed, assuming that occupations and momentum matrix elements are fixed in time (e.g. after the laser field turns off and no phenomenological dephasing is employed), then $|\sum_{\textbf{k}} \rho^{(\textbf{k})}_{cv}(t)|$ fully accounts for the time-dependence in the coherence, which makes it the ideal entity to analyze for extracting coherence lifetimes. An important point in this approach that is often not addressed is that the coherence continues to evolve even in absence of a laser field due to the occupation of CB states at different \textit{k}-points (a superposition state). Each $\rho^\textbf{k}_{cv}$ term evolves in time with the proper eigen-energy as $\sim e^{-i(\epsilon_{CB}^\textbf{k}-\epsilon_{VB}^\textbf{k}) t}$, which leads to a temporal dependence, as well as natural coherence decay even in absence of any dephasing channels in the simulation. However, this coherence decay in itself differs from traditional dephasing and arises from coherence leakage into higher modes as the superposition state delocalizes the wave functions in real-space. The resulting dynamics of $|\sum_{\textbf{k}} \rho^{(\textbf{k})}_{cv}(t)|$ is highly oscillatory, does not decay exactly to zero, and is time-linear unlike the expected exponential effect (see Fig. \ref{fig:4}(c), blue curve). As a result, we do not analyze this case, which obviously also does not cause HHG emission suppression.

In Fig. \ref{fig:4}(a-c) we analyze the BZ-averaged $\rho_{cv}(t)$ dynamics including phonons in typical laser-driven conditions. The decay towards the end of the laser is fitted to an exponentially-decaying function, yielding an effective dephasing time of $T_2^{(e-ph)}\approx 9.55$ fs. We note that this dephasing time extraction is still affected by the laser since the pulse is still `on', driving weak transitions even in the decaying part of the envelope (which ultimately increase coherences). In order to isolate the intrinsic dephasing induced by phononic DOF without effects from the laser, we perform a set of artificial simulations where the laser is abruptly and continuously halted during simulations (see Fig. \ref{fig:4}(d)). This is achieved by adding a temporal envelope $g(t)$ to the laser (see details in SI). Since the laser field is `off' after this moment in time, the results more cleanly represent coherence decay dynamics. Fitting $\rho_{cv}(t)$ after the pulse is turned off yields a value $T_2^{(e-ph)}\approx 5.69$ fs (Fig. \ref{fig:4}(e,f)). The timescales obtained here for $e$-phonon-induced dephasing under strong lasers are extremely fast, and much more rapid than those expected from $e-e$ scattering, which would indicate they are the dominant channel of decoherence in strong-field physics. 

\begin{figure*}[t]
    \centering
    \includegraphics[width=0.8\textwidth]{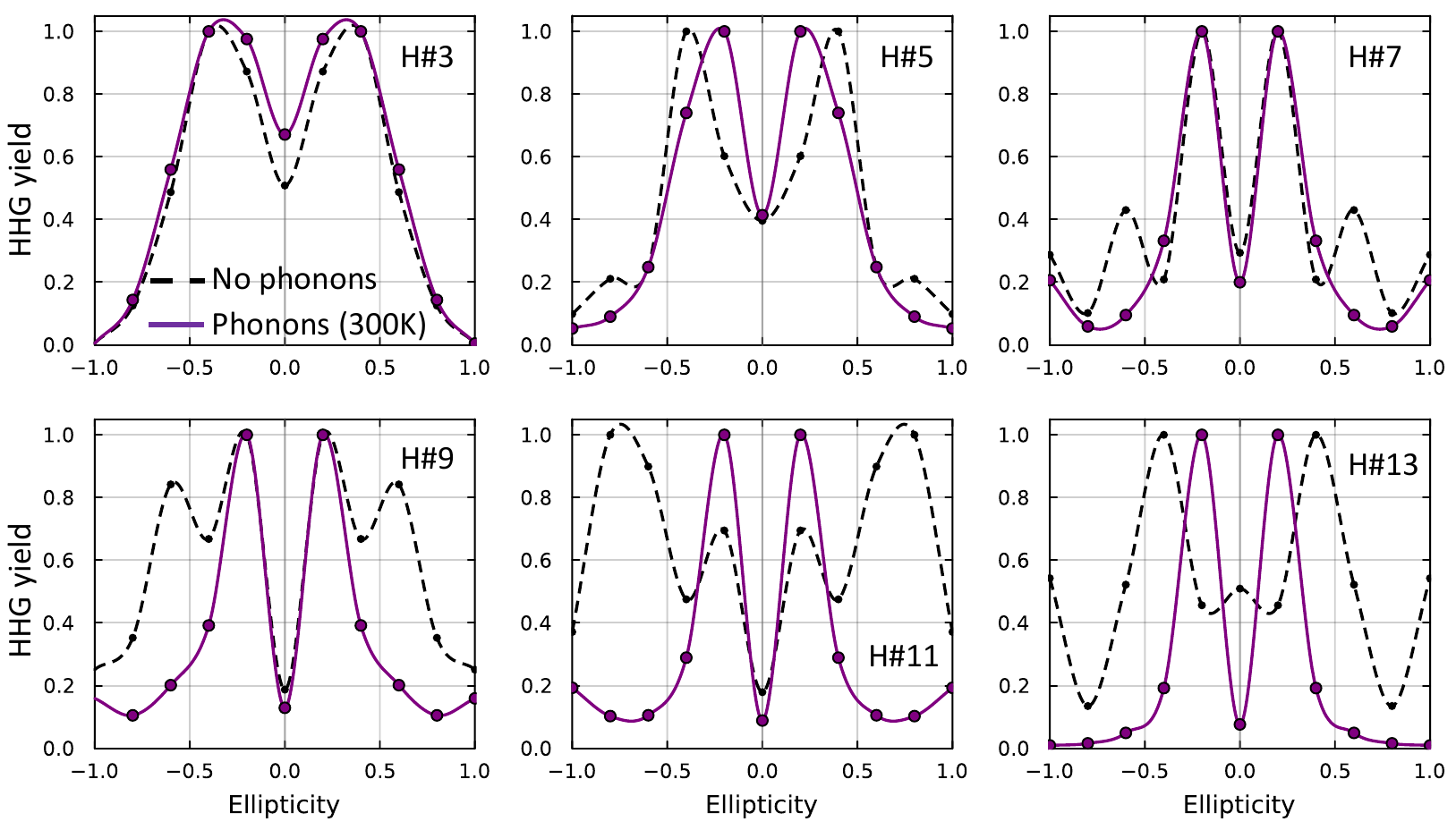}
        \captionsetup{margin=0pt}
    \caption{Ellipticity-dependence of normalized HHG yields for select harmonic orders with/without phononic interactions. Simulations performed without phenomenological dephasing.}
    \label{fig:5}
\end{figure*}

The directly extracted value of 5.69 fs is validated by a second indirect approach - we perform HHG simulations in the equilibrium case with phenomenological dephasing at this timescale, $T_2=5.69$ fs. This indeed leads to similar magnitude suppression in HHG yields (see SI). However, we note that the spectrum is not exactly reconstructed, nor the coherence dynamics, indicating that phenomenological dephasing does not have the same effect as actual inclusion of phononic DOF. From the physical standpoint of the phase scrambling mechanism, this is clear, since phenomenological dephasing suppresses $\rho_{cv}$ over time at every \textit{k}-point. On the other hand, the suppression due to phase scrambling is localized to regions in the BZ that are modified by optical phonons (e.g. the Dirac cone), and occurs instantaneously due to destructive interferences.

Lastly, we study HHG ellipticity dependence, which in graphene is well-known to maximize in $\varepsilon\neq0$ values\cite{Yoshikawa2017}. When ellipticity dependence is studied theocratically, it is often quite noisy with many side-peaks arising (see e.g. results in refs. \cite{PhysRevB.94.241107,Tancogne-Dejean2017,Baykusheva2021a,OferPRX2023}). Figure \ref{fig:5} compares HHG yields from the equilibrium case to the phononic case, showing that our simulations also find HHG yields maximize at $\varepsilon\neq0$. The phonon DOF `clean' the plots and reduce some of the minor side-peaks, better agreeing with experiments. The mechanism behind this effect is similar to the one discussed above, just that here the ellipticity tunes the HHG yield suppression. Interestingly, phononic scattering can cause minor shifts in the maximizing ellipticity value (e.g. H5 in Fig. \ref{fig:5} shifts by $\Delta\varepsilon\sim 0.2$), which could be employed for phonon occupation spectroscopy.


To summarize, we theocratically and numerically studied HHG from graphene, including optical-phonon-electron scattering under a static lattice approximation. We uncovered that phononic interactions can cause drastic HHG yield suppression, in agreement with previously unexplained experiential results. We studied the physical mechanism behind this suppression, showing it: (i) Does not occur in perturbative harmonics; (ii) Arises from interband current coupling to optical phonons; (iii) Occurs as a result of a `phononic phase scrambling' effect, where phonons induce wide phase variations in HHG that cause destructive interference; (iv) Is largely temperature-independent in graphene, but can be temperature-dependent in other systems depending on the phonon bands. We also showed that electron-phonon interactions directly couple to the interband coherence and effectively suppress coherence on timescales of $T_2\sim 5.7$ fs, meaning the ultrafast electron-phonon interactions dominate dephasing channels in strong-field physics in solids. Lastly, we explored elliptical HHG and showed that phononic interactions smoothen harmonic ellipticity dependence, and can also shift HHG maximizing ellipticities. Thus, ellipticity-dependent HHG should be employable in HHG spectroscopies of phonon occupations. 

Our results shed light on several open problems in the fields of HHG and strong-field physics. Insights obtained here can be used across material systems to both improve numerical modeling of HHG, as well as develop novel spectroscopies of electron-phonon interactions and phonon dynamics. Beyond HHG, our results should also be broadly applicable to other highly nonlinear phenomena that might couple to phonons, such as photocurrent generation\cite{Neufeld2021a,Higuchi2017,Schiffrin2013,Galler2025,Heide2021}, and Floquet phenomena and the issue of observing Floquet topological gaps in graphene\cite{Rudner2020,McIver2020,Merboldt2025,Choi2024,Wang2026}.

\vspace{20pt}

\noindent \textbf{Acknowledgments.} The authors thank Prof. Michael Kr\"uger for insightful discussions. O.N. gratefully acknowledges the Young Faculty Award from the National Quantum Science and Technology program of Israel’s Council of Higher Education Planning and Budgeting Committee and the Technion NEVET programs of the RBNI and Helen Diller Quantum Center.

\newpage

\onecolumngrid
\section{\textbf{Supplementary Information: Role of ultrafast electron-optical-phonon interactions in high harmonic generation from graphene}}

\noindent This supplementary information file contains additional technical details about simulations employed in the main text, as well as additional complementary results that support our conclusions and analysis. 

\subsection{Additional details of SBE simulations}

\noindent SBE simulations were performed by sampling the Brillouin zone using a converged \textit{k}-grid with 480$\times$480 points spanned along the directions of the reciprocal lattice vectors. We used a 4th-order Runge-Kutta scheme to solve the time evolution of the density matrix with a converged time step of 0.2 a.u. Simulations employed the following `super-sine' form for the laser envelope function\cite{OferRingcurrentPRL2019}:
\begin{equation}
f(t)=\sin{\left( \frac {\pi t}{T_{pulse}} \right)}^{\pi\left|t/T_{pulse}-0.5\right|/\sigma}
\end{equation}
, with $\sigma=0.75$ and $T_{pulse}$ being the total pulse duration. In the decay analysis of $\rho_{cv}(t)$ in Fig. 4 in the main text, the additional smooth envelope function started at $t_{cut}=5 \frac{2\pi}{\omega}$, and had a decay time of $\tau=\frac{1}{2}\frac{2\pi}{\omega}$, with the following functional form:
\begin{equation}
\begin{cases} 
g(t)=1 & t\leq t_{cut}
\\
g(t)=\exp\left( 1-\frac{1}{1- (\frac{t-t_{cut}}{\tau})^2}\right) & t_{cut}<t< t_{cut}+\tau 
\\
g(t)=0 & t_{cut}+\tau \leq t
\end{cases}
\end{equation}
\noindent , where both $\tau$ and $t_{cut}$ are synced to moments in time where the electric field vanishes.

\noindent We employed a lattice parameter for graphene at the experimental values of $a=2.46 $ \AA, with hopping parameters, $t_1=2.72$ eV, $t_2= 0.3$ eV. In all simulations we softened the graphene singularity of the transition dipole matrix elements at the Dirac cone to a level of $\frac{\eta}{t_1}=10^{-3}$, where $\eta$ addressed the divergence. 

For simulations including phononic DOF, the hopping amplitudes were varied as: $\tilde{t_1}(\textbf{R}_i)=t_1\exp\left( -3\left(\frac{ \left\vert \textbf{R}_i\right\vert}{R_0}-1 \right) \right)$ where $\textbf{R}_i$ are the displaced lattice nearest neighbor distances and $R_0$ is the equilibrium nearest neighbor bond length of $1.42$ \AA. The sampling of thermal perturbations in the phononic simulations was performed by drawing a Gaussian random variable over both Cartesian displacement components, accounting for the two $\Gamma$-point optical phonon polarizations (LO and TO). Each component was sampled independently with a standard deviation determined from the thermal variance of the displacement. We treated the phonons as a quantum harmonic oscillator of the form $H_{ph}=\frac{1}{2}\left( P^2 + \omega^2 Q^2 \right)$, with $P$ the phonon canonical momenta and $Q$ the canonical position coordinate. The eigenmodes of $H_{ph}$ were occupied according to Bose–Einstein statistics, such that the canonical position satisfies:
\begin{equation}
\langle Q^2 \rangle = \frac{\hbar}{2\omega}\coth\left( \frac{\hbar\omega}{2k_BT}\right).
\end{equation}

\noindent For the $\Gamma$ optical phonon, $Q$ is related to the atomic displacements via the normalized optical phonon eigenvectors (assuming equal carbon atom masses on both A/B sublattice sites):
\begin{equation}
\mathbf{u}_A = \frac{1}{\sqrt{2M_C}}\mathbf{Q} ; \qquad
\mathbf{u}_B = -\frac{1}{\sqrt{2M_C}}\mathbf{Q}
\end{equation}

\noindent where $M_C$ is the mass of a carbon atom. The relative atomic displacement in the lattice is therefore $\boldsymbol{\Delta} = \mathbf{u}_B - \mathbf{u}_A = -\sqrt{\frac{2}{M_C}}\,\mathbf{Q}$. This yields, per cartesian component of the displaced atom, the following relations:
\begin{equation}
\langle \Delta_i^2 \rangle
=
\frac{2}{M_C}\langle Q_i^2 \rangle
=
\frac{\hbar}{M_C\omega_{ph}}
\coth\left( \frac{\hbar\omega_{ph}}{2k_BT}\right).
\end{equation}
\noindent Using this relation between atomic displacement and temperature, and the above described random sapling procedure of the distribution functions, individual `snapshots' of the thermally occupied lattice were drawn and the HHG spectra was simulated from each snapshot individually. The coherently summed emission from all snapshots provides HHG simulations including interactions with these phononic DOF. 

The Gaussian focal beam intensity averaging procedure presented in Fig. 2(f) in the main text employed the following procedure. First, we calculated HHG spectra including full phononic snapshot coherent averaging for different laser peak powers, sampling the gaussian beam power distribution that is described by $I(r)=I_0\exp(-2(r/w)^2)$, with $I_0$ the beam peak power, $w$ the beam waist, and $r$ a radial coordinate away from the center. We performed $7$ simulations at power ranging from $0.85I_0$ up to $I_0$ with $r_i$ sampled equidistantly. Each of the obtained HHG spectra was then incoherently summed with proper weights based on the annulus area between the corresponding radii and the next larger equidistant radius. 

\subsection{Additional results}

\noindent Here we add additional complementary results to those presented in the main text. First, Fig. S1 presents convergence data of HHG spectra in typical conditions in the phononic case vs the number of snapshots employed for sampling the optical phonon distribution (i.e. $N_{snap}$). Most HHG spectra were seen to converge after relatively few snapshots (order $N_{snap}\sim300$). However, in some laser conditions we noted slightly more snapshots were needed. Full convergence is typically obtained at $N_{snap}=650$ regardless of the laser regime (see Fig. S1). All data in the main text employed a stricter criterion of $N_{snap}=750$ for phononic cases.

\begin{figure*}[h]
    \centering
    \includegraphics[width=0.5\textwidth]{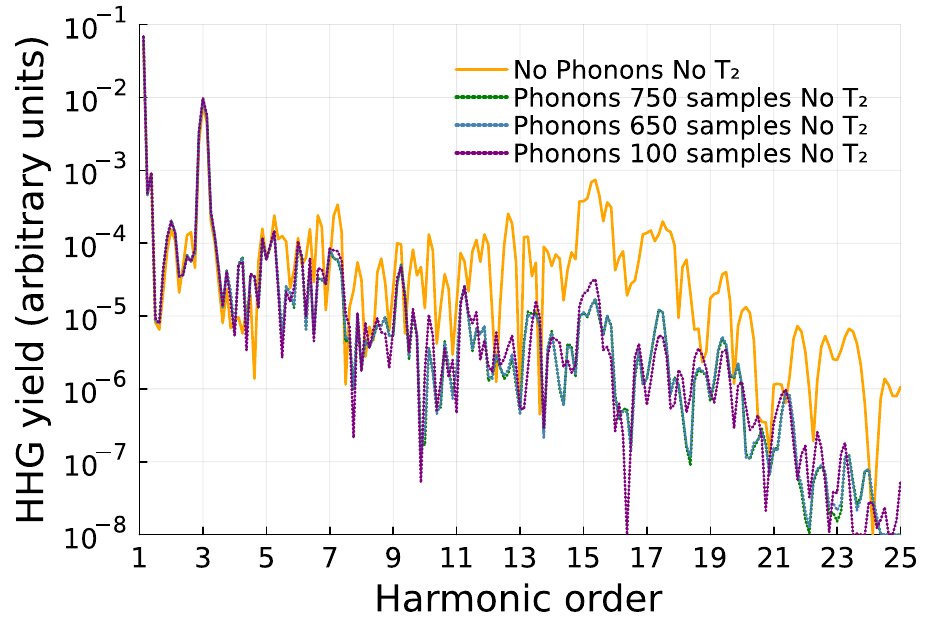}
    \captionsetup{margin=0pt}

    \caption{Convergence of HHG emission in the phononic case with respect to number of phononic configuration sampling, $N_{snap}$. Simulations performed in similar conditions to Fig. 2 in main text at 300K.}
    \label{fig:S1}
\end{figure*}

Figure S2 complements results in Fig. 2(d) in the main text, showing HHG emission in the phononic case with a very high temperature of 1500K and 2500K. Under such conditions, substantial occupation of the optical phonons is obtained, which starts impacting the HHG emission characteristics. Therefore, in the case of graphene, we only expect temperature-dependent HHG effects to start appearing in very high temperatures (where very low temperatures are dominated by zero point motion, as discussed in the main text). However, we also note that in practice the laser pulses are likely to substantially occupy optical phonon modes during the electron-driven dynamics due to direct pumping as well as indirect energy transfer, which might lead to conditions in experiments being closer to the high-temperature conditions simulated here. Moreover, under such high temperatures other phonon modes beyond $\Gamma$ and acoustic should also contribute to the dynamics. 

\begin{figure*}[h!]
    \centering
    \includegraphics[width=0.5\textwidth]{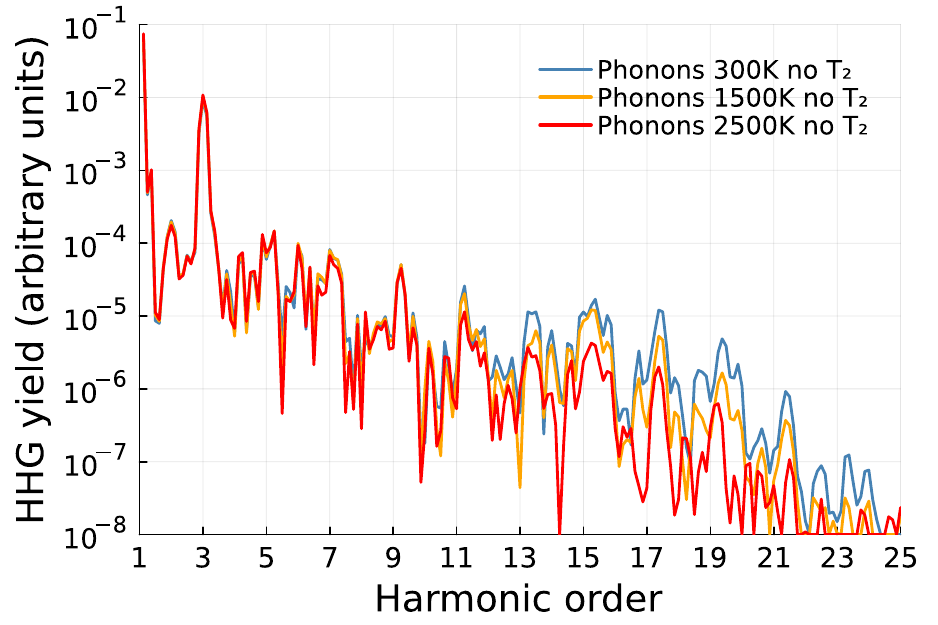}
    \captionsetup{margin=0pt}

    \caption{HHG temperature dependence from graphene in high temperature conditions. Simulations performed in similar conditions to Fig. 2(d) in main text.}
    \label{fig:S2}
\end{figure*}
\FloatBarrier

Next, we present in Fig. S3 all harmonic phase distributions across snapshots for the interband HHG channel (with intraband emission fixed at the equilibrium case). This complements Fig. 3(c-d) in the main text that only presented select harmonic orders. The phase distributions are very wide for all harmonics above 5'th order, promoting destructive interferences, as discussed in the main text.

\begin{figure*}[h!]
    \centering
    \includegraphics[width=0.7\textwidth]{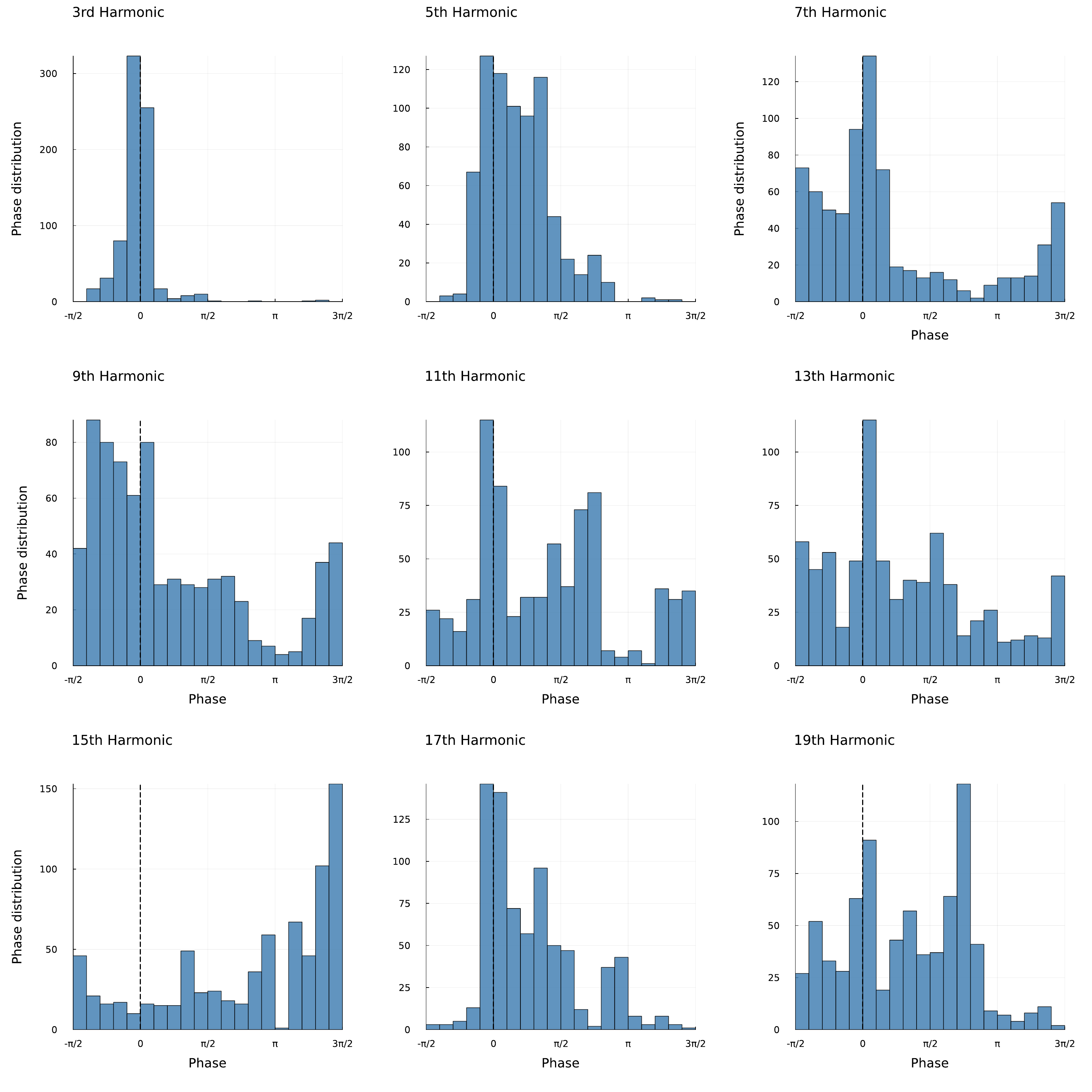}
    \captionsetup{margin=0pt}

    \caption{Phononic phase scrambling effect in interband HHG. Histograms of harmonic phases across snapshots are presented for all odd harmonic orders, complementing select harmonic orders presented in the main text in Fig. 3(c-d) (identical simulation methodology and conditions are employed here).}
    \label{fig:S3}
\end{figure*}
\FloatBarrier

Lastly, Fig. S4 presents HHG yields in similar conditions to Fig. 1(a,b) in the main text (equilibrium case), but with even shorter phenomenological dephasing times of $T_2=5.69$ fs for the equilibrium case (which is precisely the timescale obtained independently from fitting the interband coherence in the phononic case in Fig. 4(f) in the main text). This yields a very substantial suppression of HHG yields that mimics the results of the simulation that includes phononic DOF. This result therefore constitutes a second independent (and indirect) approach that validates this dephasing timescale in graphene induced by ultrafast electron-optical-phonon interactions. Notably, the spectrum is not exactly reconstructed, meaning such phenomenological terms only roughly mimic yield suppression (see discussion in the main text).  

\begin{figure*}[h!]
    \centering
    \includegraphics[width=0.5\textwidth]{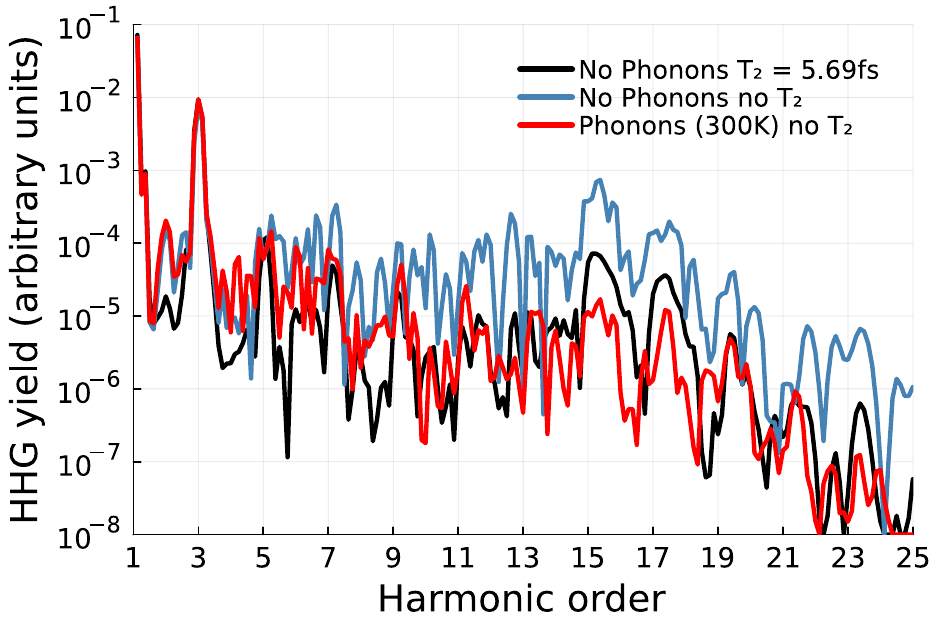}
    \captionsetup{margin=0pt}

    \caption{HHG yields comparing the phononic case with the equilibrium case with phenomenological dephasing set to $T_2=5.69$ fs, which yields similar magnitudes of HHG suppression. Simulations performed in similar conditions to Fig. 1(a,b) in the main text.}
    \label{fig:S4}
\end{figure*}
\FloatBarrier

\bibliography{references}

\end{document}